# Conceptualizing and Evaluating Replication Across Domains of Behavioral Research

Jennifer L. Tackett & Blakeley B. McShane
Northwestern University

**Abstract:**

We discuss the authors' conceptualization of replication, in particular the false dichotomy of direct versus conceptual replication intrinsic to it, and suggest a broader one that better generalizes to other domains of psychological research. We also discuss their approach to the evaluation of replication results and suggest moving beyond their dichotomous statistical paradigms and employing hierarchical / meta-analytic statistical models.

**Main Text:**

We thank Zwaan *et al*. (2018; hereafter ZELD) for their review paper on replication and strongly endorse their call to make replication mainstream. Nonetheless, we find their conceptualization of and recommendations for replication problematic.

Intrinsic to ZELD's conceptualization is a false dichotomy of direct versus conceptual replication, with the former defined as "a study that attempts to recreate the critical elements (e.g., samples, procedures, and measures) of an original study" (p. 8) and the latter as a "study where there are changes to the original procedures that might make a difference with regard to the observed effect size" (p. 9). We see problems with both of ZELD's definitions and the sharp dichotomization intrinsic to their conceptualization.

In terms of definitions, first, ZELD punt in defining direct replications by leaving unspecified the crucial matter of what constitutes the "critical elements of an original study." Specifying these is nontrivial if not impossible in general and likely controversial in specific. Second, they are overly broad in defining conceptual replications: under their definition, practically all behavioral research replication studies would be considered conceptual. To understand why, consider large-scale replication projects such as the Many Labs project (Klein *et al*., 2014) and Registered Replication Reports (RRRs; Simons *et al*., 2014) where careful measures were taken such that protocols were followed identically across labs in order to achieve near exact or direct replication; even under such strict conditions, effect sizes did indeed differ (i.e., are heterogeneous or contextually variable) across labs and to roughly the same degree as sampling variation (McShane, *et al*., 2016; Stanley *et al*., 2017). This renders ZELD's suggestion of conducting direct replication infeasible: even if defining the "critical elements" were possible, recreating them in a manner that maintains the effect size homogeneity they insist on for direct replication seems impossible.

In addition, and also due to these Many Labs and RRR results, the sharp dichotomization of direct versus conceptual replication intrinsic to ZELD's conceptualization is unrealistic in practice. Further, even were it not, replication designs with hybrid elements (e.g., where the theoretical level is "directly" replicated but the operationalization is systematically varied) are an important future direction—particularly for large-scale replication projects (Tackett, *et al.*, 2017b)—not covered by ZELD's conceptualization.

Instead, and in line with ZELD's mention of "extensions," we would like to see a broader approach to conceptualizing replication, and in particular one that better generalizes to other domains of psychological research. Specifically, large-scale replications are typically only possible when data collection is fast and not particularly costly, and thus they are, practically speaking, constrained to certain domains of psychology (e.g., cognitive and social). Consequently, we know much less about the replicability of findings in other domains (e.g., clinical and developmental) let alone how to operationalize replicability in them (Tackett *et al.*, 2017a; Tackett, Brandes, & Reardon, 2018). In these other domains, where data collection is slow and costly but individual datasets are typically much richer, we recommend that rather than taking a prospective approach to replication, a retrospective approach that leverages the large amount of shareable archival data across sites is not only valuable but also potentially preferable (Tackett *et al.*, 2017b; Tackett *et al.*, 2018).

This will require not only a change in both infrastructure and incentive structures but also a better understanding of appropriate statistical approaches for analyzing pooled data (i.e., hierarchical models) and more complex effects (e.g., curve or function estimates as opposed to point estimates); lab-specific moderators most relevant to include in such analyses; additional method factors that drive heterogeneity (e.g., drop out mechanisms in longitudinal studies); and how to harmonize measurements across labs (e.g., if they use different measures of depression).

It may also require an overhaul of what constitutes a replication. ZELD suggest three ways of statistically evaluating a replication, all of which are based on the null hypothesis significance testing (NHST) paradigm and the dichotomous *p*-value thresholds intrinsic to it. Such thresholds, whether in the form of *p*-values or other statistical measures such as confidence intervals and Bayes Factors, (i) lead to erroneous reasoning (McShane and Gal, 2016, 2017), (ii) are a form of statistical alchemy that falsely promise to transmute randomness into certainty (Gelman, 2016) thereby permitting dichotomous declarations of truth or falsity, binary statements about there being "an effect" or "no effect," a "successful replication" or a "failed replication", and (iii) should be abandoned (McShane *et al.*, 2017; McShane and Gelman, 2017).

Instead, we would like to see replication efforts statistically evaluated via hierarchical / meta-analytic statistical models. Such models can directly estimate and account for contextual variability (i.e, heterogeneity) in replication efforts, which is critically important given, as per the Many Labs and RRR results, that such variability is roughly comparable to sampling variability even when explicit efforts are taken to minimize it and is typically many times larger in more standard sets of studies when they are not (van Erp, *et al.*, 2017; Stanley *et al.*, 2017). Importantly, they can also account for differences in methods factors such as dependent variables, moderators, and study designs (McShane and Bockenholt, 2017a, 2017b) and for varying treatment effects (Gelman, 2015) thereby allowing for a much richer characterization of a research domain and application to the hybrid replication designs discussed above. We would also like to see the estimates from these models considered alongside additional factors such as prior and related evidence, plausibility of mechanism, study design and data quality in order to provide a more holistic evaluation of replication efforts.

Our suggestions for replication conceptualization and evaluation forsake the false promise of certainty offered by the dichotomous approaches favored by the field and by ZELD. Consequently, they will seldom if ever deem a replication effort a "success" or a "failure" and indeed reasonable people following them may disagree about the degree of replication success. However, by accepting uncertainty and embracing variation (Carlin, 2016; Gelman, 2016), we believe these suggestions will help us learn much more about the world.

**References:**

Carlin, J. B. (2016). Is reform possible without a paradigm shift? The American Statistician, supplemental material to the ASA statement on p-values and statistical significance 10.

Gelman, A. (2015). The connection between varying treatment effects and the crisis of unreplicable research: A bayesian perspective. Journal of Management 41, 2, 632–643.

Gelman, A. (2016). The problems with p-values are not just with p-values. The American Statistician, supplemental material to the ASA statement on p-values and statistical significance 10.

Klein, R. A., Ratliff, K., Nosek, B. A., Vianello, M., Pilati, R., Devos, T., Galliani, E. M., Brandt, M., van 't Veer, A., Rutchick, A. M., Schmidt, K., Bahnik, S., Vranka, M., IJzerman, H., Hasselman, F., Joy-Gaba, J., Chandler, J. J., Vaughn, L. A., Brumbaugh, C., van swol, L., Wichman, A., Packard, G., Brooks, B., Cemalcilar, Z., Storbeck, J., Bocian, K., Levitan, C., Bernstein, M. J., Krueger, L. E., Eisner, M., Davis, W. E., Nier, J. A., Nelson, A. J., Steiner, T. G., Mallett, R., Thompson, D., Huntsinger, J. R., Morris, W., Skorinko, J., and Kappes, H. (2014). Investigating variation in replicability: A "many labs" replication project. Social Psychology 45, 3, 142–152.

McShane, B. B. and Bockenholt, U. (2017). Single paper meta-analysis: Benefits for study summary, theory-testing, and replicability. Journal of Consumer Research 43, 6, 1048– 1063.

McShane, B. B. and Bockenholt, U. (2017; forthcoming). Multilevel multivariate meta- analysis with application to choice overload. Psychometrika .

McShane, B. B., Bockenholt, U., and Hansen, K. T. (2016). Adjusting for publication bias in meta-analysis: An evaluation of selection methods and some cautionary notes. Perspectives on Psychological Science 11, 5, 730–749.

McShane, B. B. and Gal, D. (2016). Blinding us to the obvious? the effect of statistical training on the evaluation of evidence. Management Science 62, 6, 1707–1718.

McShane, B. B. and Gal, D. (2017). Statistical significance and the dichotomization of evidence. Journal of the American Statistical Association, 112(519), 885-895.

McShane, B. B., Gelman, A. (2017). Abandon statistical significance. Nature 551(7682), 558.

McShane, B. B., Gal, D., Gelman, A., Robert, C., and Tackett, J. L. (2017). Abandon statistical significance. Tech. rep., Northwestern University.

Open Science Collaboration (2015). Estimating the reproducibility of psychological science. Science 349, 6251, aac4716.

Simons, D. J., Holcombe, A. O., and Spellman, B. A. (2014). An introduction to registered replication reports at perspectives on psychological science. Perspectives on Psychological Science 9, 5, 552–555.

Stanley, T. D., Carter, E. C., Doucouliagos, H. (2017). What meta-analyses reveal about the replicability of psychological research. Deakin Laboratory for the Meta-Analysis of Research, Working Paper, November 2017.

Tackett, J. L., Brandes, C. M., Reardon, K. W. (2018). Leveraging the Open Science Framework in clinical psychological assessment research. Manuscript in press, *Psychological Assessment*.

Tackett, J.L., Lilienfeld, S.O., Patrick, C.J., Johnson, S.L, Krueger, R.F, Miller, J.D., Oltmans, T.F, and Shrout, P.E. (2017a). It's time to broaden the replicability conversation: Thoughts for and from clinical psychological science. *Perspectives on Psychological Science* 12:5, 742-756.

Tackett, J. L., McShane, B. B., Bockenholt, U., and Gelman, A. (2017b). Large scale replication projects in contemporary psychological research. Tech. rep., Northwestern University.

van Erp, S., Verhagen, A. J., Grasman, R. P. P. P., & Wagenmakers, E.-J. (2017). Estimates of between-study heterogeneity for 705 meta-analyses reported in Psychological Bulletin from 1990-2013. Journal of Open Psychology Data, in press.

Zwaan, R. A., Etz, A., Lucas, R. E., Donnellan, B. (2018). Making replication mainstream. Behavioral and Brain Sciences.